\documentclass[aps,prl,twocolumn,superscriptaddress]{revtex4}
\usepackage{epsfig}

\newcommand{\cecoin}{Ce$_x$La$_{1-x}$CoIn$_5$}
\newcommand{\celapb}{Ce$_x$La$_{1-x}$Pb$_3$}

\begin{document}

\title {
Strongly Inhomogeneous Phases and Non-Fermi Liquid Behavior \\
in Randomly Depleted Kondo Lattices
}

\author{Ribhu K. Kaul}
\affiliation{Department of Physics, Duke University, Durham, NC 27708-0305, USA}
\affiliation{\mbox{Institut f\"ur Theorie der Kondensierten Materie,
Universit\"at Karlsruhe, 76128 Karlsruhe, Germany}}
\author{Matthias Vojta}
\affiliation{\mbox{Institut f\"ur Theorie der Kondensierten Materie,
Universit\"at Karlsruhe, 76128 Karlsruhe, Germany}}
\date{\today}

\begin{abstract}
We investigate the low-temperature behavior of Kondo lattices
upon random depletion of the local $f$-moments, by using
strong-coupling arguments and solving SU($N$) saddle-point equations on large lattices.
For a large range of intermediate doping levels,
between the coherent Fermi liquid of the dense lattice and the single-impurity
Fermi liquid of the dilute limit, we find strongly inhomogeneous states that
exhibit distinct non-Fermi liquid characteristics.
In particular, the interplay of dopant disorder and strong interactions leads
to rare weakly screened moments which dominate the bulk susceptibility.
Our results are relevant to compounds like \cecoin\ and \celapb.
\end{abstract}

\maketitle


Heavy-fermion metals \cite{hewson} are in the focus of correlated electron
physics for numerous reasons. They feature a strongly correlated heavy Fermi-liquid (HFL) state
that arises from the Kondo screening of a lattice of localized $f$-moments
by conduction electrons.
The HFL phase can undergo quantum phase transitions to various magnetic phases,
e.g., spin-density wave and spin-glass states.
Remarkably, even the Fermi-liquid (FL) regime of heavy-fermion metals is not completely understood.
Experimentally, different observables such as entropy, bulk susceptibility, and resistivity,
are employed to define characteristic crossover temperatures,
but their precise relation to microscopic processes is not known.
Usually, the behavior cannot be characterized by a single energy
scale only;
this situation is markedly different from that of a single magnetic impurity
in a metal, where the single-impurity Kondo temperature $T_K^{(1)}$
is the only low-energy scale, and all thermodynamic observables are universal
functions of $T/T_K^{(1)}$ only.

A fascinating modification of Kondo lattices is the depletion of the
local-moment sector, by randomly replacing magnetic by non-magnetic
rare-earth ions.
This allows one to continuously tune the system between the dense Kondo lattice
and the isolated impurity -- both limits are
known to exhibit FL behavior
(provided that no magnetic or superconducting order intervenes).
This raises fundamental questions:
(i) Do the two FL phases evolve continuously into each other upon changing
the concentration of $f$-moments?
(ii) Over what range of doping is a picture of independent impurities applicable?

\begin{figure}[!t]
\centerline{\epsfig{file=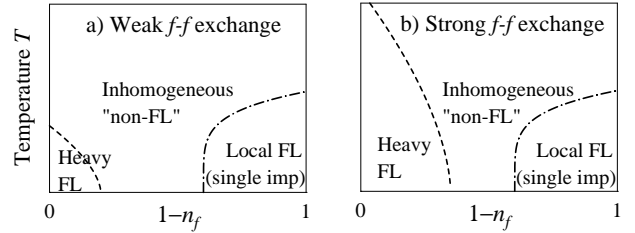,width=3.4in}}
\caption{
Schematic phase diagram of the depleted Kondo lattice.
In the dense limit, $n_f \lesssim 1$, a coherent heavy FL is formed below $T_{\rm coh}$ (dashed),
whereas in the dilute limit, $n_f\to0$, single-impurity physics prevails, with local FL behavior
below $T_K^{(1)}$ (dash-dot).
There is a large regime of intermediate $n_f$ where low-energy properties are strongly
inhomogeneous, and FL behavior is suppressed.
a) For weak inter-moment exchange we have $T_{\rm coh} < T_K^{(1)}$,
whereas b) stronger inter-moment exchange enhances $T_{\rm coh}$ and
shifts the boundary of the inhomogeneous regime towards smaller $n_f$.
}
\label{fig:pd}
\end{figure}

Early experimental studies on \celapb\ \cite{celapb} indicated that a single-impurity
picture may be valid over a rather large range of $f$-moment concentration:
it was found that even at $x=0.6$ both transport and thermodynamic measurements
could be described using a single-impurity Kondo model down to the lowest temperatures
studied (one-tenth of the estimated $T_ K^{(1)}$).
These results are particularly surprising because the ``Kondo screening clouds''
should be strongly overlapping in this regime.
In contrast, recent detailed investigations of \cecoin\ \cite{naka1}
interestingly revealed different properties:
FL behavior, as inferred from the resistivity, is
quickly suppressed to very low $T$ when moving away both from the dense and
dilute limit.
The energy scale characterizing the onset of moment screening in the
susceptibility is enhanced by more than an order of magnitude in the
dense case compared to the dilute limit.

On the theoretical side,
the behavior of depleted Kondo lattices has been studied rather little.
The experiments on \cecoin\ triggered an interpretation in terms
of a phenomenological two-fluid model \cite{naka2}
(see Ref.~\onlinecite{barzy} for a related mean-field approach).
However, detailed microscopic investigations taking into account the dopant disorder
(apart from averaged treatments using CPA \cite{cpa}) are lacking.

The aim of this Letter is to close this gap:
using both strong-coupling arguments and detailed numerical results,
obtained from solving SU($N$) saddle-point equations on finite lattices,
we shall present a consistent picture of the evolution of the HFL upon
random $f$-moment depletion ~\cite{fn1,fn1b}.

Our main findings can be summarized as follows.
The crossover from HFL to the single-impurity FL takes place
at a concentration of $f$ moments that strongly depends on
the number of delocalized electrons per unit cell, $n_c$.
In particular, at large $n_c$
single-impurity behavior is observed even for dense $f$
moment concentrations (as in \celapb).
Between the two FLs we find a wide crossover regime
with strongly inhomogeneous states at low temperature:
The local density of states (LDOS) in the conduction band near the Fermi level
shows signatures of localization, associated with island formation,
and the uniform susceptibility does not saturate down to very low temperatures
due to the presence of poorly screened moments.
Importantly, the broad distributions of local quantities, such as
the local susceptibility $\chi_{\rm loc}$, and the associated non-FL behavior do not
originate from disorder in the {\em bare} LDOS or in the local Kondo couplings
(as in so-called Kondo disorder models \cite{KondoDis}),
but arise through the collective screening
of the randomly positioned $f$-moments by the conduction electrons.
Our results are summarized in the phase diagrams, Fig.~\ref{fig:pd}.



{\it Model.}
The standard Kondo lattice model with randomly depleted $f$-moments reads
\begin{eqnarray}
  H_{\rm KLM} &=& \sum_k \epsilon_k c^{\dagger}_{k\alpha} c_{k\alpha} +
  \frac{J_K}{2}\sum_{\{r\}}
  \vec S_r \cdot c^{\dagger}_{r\alpha} \vec \sigma_{\alpha\alpha'} c_{r\alpha'},
\label{KH}
\end{eqnarray}
where $c_{k\alpha}$ represent the conduction electrons with momentum $k$, spin
$\alpha$, a band width $D$, and a band filling of $n_c$ ($n_c = 1$ corresponds to half filling,
in the following we shall assume $n_c<1$).
The $\vec S_r$ are the spin-$1/2$ moments on the $N_f$ {\em occupied} sites $\{r\}$
of the $f$ electron lattice, with a concentration of $n_f = N_f/N_s$ where $N_s$ is
the number of conduction electron sites ($n_f=1$ is the dense Kondo lattice).


{\it Strong-coupling analysis.}
In the limit $J_K \gg t$ in Eq.~(\ref{KH})
the $c$-electrons and $f$-spins form local singlets.
In the dilute limit, $n_f \ll 1$, one ends up with $N_f$ local singlets
which act as potential scatterers for the remaining $c$-electrons -- the system
is a FL \cite{fn1,fn1b}.
In the dense limit, $n_f=1$, it is useful to think in terms of holes:
after forming $N_s$ singlets we are left with $N_s(1-n_c)$ mobile holes in the conduction
band which are HFL quasi-particles: clearly this limit is also a FL.
Now, as the $f$-moments are randomly depleted, the band of HFL quasi-particles is also
depleted.
Let us focus on $n_f=n_c$: In the strong-coupling limit all conduction electrons are
bound into localized singlets, and the system has a gap of order $J_K$ to all
excitations -- this is apparently {\em not} a Fermi liquid.
This discussion implies that the ground state of Eq.~(\ref{KH})
evolves discontinuously upon tuning of $n_f$.
In particular, $n_f = n_c$ appears to be the dividing line
between HFL and single-impurity FL behavior.
In the region $n_f \approx n_c$, we expect that the system is most sensitive
to randomness (introduced by $f$-spin depletion), such that
localization phenomena and strong inhomogeneities are likely to occur.
These general arguments will be supported by our mean-field calculations below.
We note that the gap at $n_f=n_c$ (of similar origin as the gap of the Kondo insulator, $n_f=n_c=1$)
is a feature of the strong-coupling limit, and not expected to survive
for $J_K \lesssim t$.


\begin{figure}[!t]
\centerline{\epsfig{file=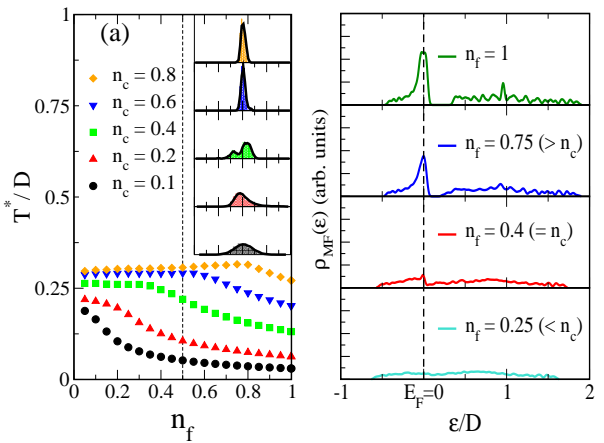,width=3.1in}}
\caption{
(color online)
Summary of the mean-field results:
(a) $T^*\!=\!\langle b_r^2\rangle/D$ as function of the local-moment
concentration $n_f$ for $J_K/D\!=\!1.25$ and different band fillings $n_c$.
Note the sharp change around $n_c\!=\!n_f$. The inset shows
the distribution of $b_r^2/\langle b_r^2 \rangle$ for the different $n_c$ at fixed $n_f\!=\!0.5$.
The strongly inhomogeneous regime is characterized by a
bimodal distribution, here most pronounced at $n_c\!=\!0.4$.
(b) Evolution (from top to bottom) of the mean-field density of states from the HFL
to the dilute impurity limit at fixed $n_c\!=\!0.4$.
}
\label{fig:b}
\end{figure}

{\it Mean-field theory.}
To obtain quantitative results, we now turn to a solvable large-$N$ limit of Eq.~(\ref{KH})
that allows us to access arbitrary $J_K/t$.
Using slave fermions to represent the local moments,
$\vec S_r = \frac{1}{2}
f^{\dagger}_{r\alpha} \vec \sigma_{\alpha\alpha'} f_{r\alpha'}$
with $f^{\dagger}_{r\alpha} f_{r\alpha}=1$,
decoupling the Kondo interactions with auxiliary fields $b_r$,
and neglecting temporal fluctuations of the $b_r$,
we obtain the mean-field Hamiltonian
\begin{eqnarray}\label{mf1}
  H_{\rm mf} & = & \sum_k \epsilon_k c^{\dagger}_{k\alpha} c_{k\alpha} \\
  &+&  \sum_{\{r\}} \mu_r f^{\dagger}_{r\alpha}f_{r\alpha} -
\sum_{\{r\}} \left( b_r c^{\dagger}_{r\alpha}
  f_{r\alpha} + \mbox{h.c.} \right).
\nonumber
\end{eqnarray}
Note that $H_{\rm mf}$ represents the $N=\infty$ saddle-point solution
of the SU($N$) Kondo lattice model, with a fully antisymmetric representation of
the local moments.
The mean-field parameters $b_r, \mu_r$ are determined by the self-consistency conditions
\begin{equation}
\label{selfcon}
1     = \langle f^{\dagger}_{r\alpha} f_{r\alpha} \rangle  \,,~
2 b_r =  J_K \langle c^{\dagger}_{r\alpha} f_{r\alpha} \rangle.
\end{equation}

At high temperatures the only solution is $b_r=\mu_r=0$, corresponding to
quasi-free local moments with Curie susceptibility.
Upon lowering $T$, all $b_r$ will become non-zero at $T_K^{(1)}$,
the (mean-field) single-impurity Kondo temperature,
which is determined by $J_K$ and the LDOS of the conduction band only.
Hence, the onset of Kondo screening at $T_K^{(1)}$ \cite{fn4}
does not depend on $n_f$ and on the particular disorder realization.
Within mean-field theory, the simplest energy scale describing the {\em low-temperature}
behavior is $T^{*}=\langle |b_r(T\!=\!0)|^2\rangle/D$,
where $\langle...\rangle$ denotes a disorder average.
For a single impurity, the ratio $T^{*}/T_K^{(1)}$ is a constant of order unity.
In contrast, in the dense Kondo lattice $T^{*}/T_K^{(1)} \ll 1$
depends on the band filling $n_c$ (but not on $J_K$ at weak coupling);
$T^{*}$ can be associated with the coherence temperature $T_{\rm coh}$ below which
the Kondo lattice behaves as a FL~\cite{burdin,pruschke}.


{\it Numerical results.}
We have performed large-scale numerical simulations of the
mean-field equations on square lattices of up to $20\times20$ sites
for different $n_c$ and $n_f$ \cite{fn3}.
Fig.~\ref{fig:b}(a) shows a summary of the mean-field condensation parameters.
Upon increasing $n_f$ the scale $T^*$ stays roughly constant until $n_f \approx n_c$
and then it drops sharply towards its $n_f=1$ value.
This supports the picture that emerged from strong-coupling: $n_f \approx n_c$
separates HFL from single-impurity FL behavior.
The distribution function of $\langle b_r^2\rangle$ becomes bimodal
when $n_f \approx n_c$ pointing to strong inhomogeneities at this filling.
Fig.~\ref{fig:b}(b) shows the evolution of the mean-field density of states of the HFL upon random
$f$-moment dilution.
For the dense lattice the Fermi level lies in region of the lower band with a large density of states.
The number of states in this heavy-electron band is reduced upon $f$-moment depletion.
Near $n_f=n_c$ the two bands merge, and the Fermi level lies near the edge of the
bands where the states are most localized.
(We have studied the inverse participation ratio of the mean-field wave functions~\cite{TBP}
and confirmed this expectation.)
The system at this filling is thus in a strongly inhomogeneous phase.

\begin{figure}[!t]
\centerline{\epsfig{file=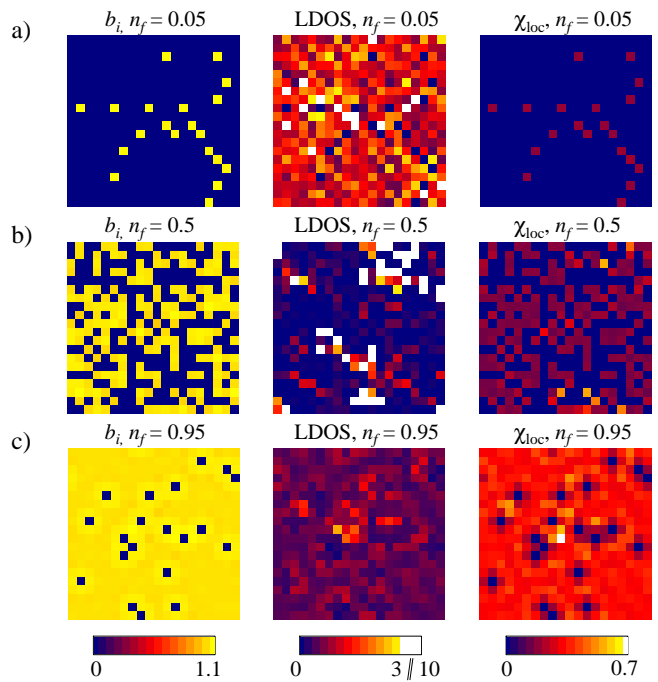,width=3.6in}}
\caption{
(color online) Spatial distribution of the slave boson amplitudes ($b_r$, left),
the low-energy LDOS ($A_r(\omega=0)$, middle),
and the local susceptibilities ($\chi_{{\rm loc,}r}(T\!=\!10^{-3})$, right)
for different $n_f=0.05,0.5,0.95$
at a band filling of $n_c=0.8$.
The Kondo coupling is $J_K/D=1.25$.
}
\label{fig:ldos_nf}
\end{figure}

\begin{figure}[!b]
\centerline{\epsfig{file=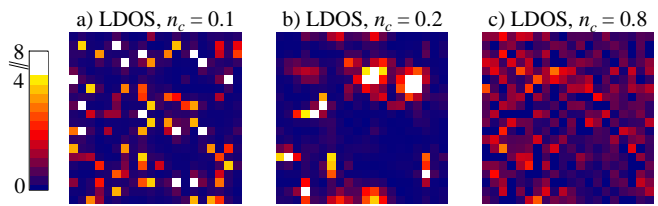,width=3.6in}}
\caption{
(color online) Spatial distribution of the LDOS, $A_r(\omega=0)$,
for a specific disorder realization with $n_f=0.2$,
but for varying conduction electron filling $n_c=0.1,0.2,0.8$.
The crossover from the single-vacancy (a) to the single-impurity (c) regime
can be nicely seen, with a tendency towards island formation in between.
The $f$-moment locations can be identified from the dark sites in (c).
}
\label{fig:ldos_nc}
\end{figure}

To gain a better understanding of the underlying microscopics, let us
focus on local properties of a particular disorder realization.
Fig.~\ref{fig:ldos_nf} shows the spatial distributions of
the local magnetic susceptibility $\chi_{\rm loc}$ and
the LDOS $A(\omega)$ (defined as the imaginary part of the local
$c$ electron propagator) at the Fermi level
(averaged over a small energy interval to avoid discretization effects).
Let us first discuss the limiting cases:
in the dilute limit, $n_f \ll 1$,
we expect a single-impurity picture to hold:
Each moment is individually Kondo-screened, and the LDOS
is suppressed at the {\em $f$-moment locations}
because of the resonant scattering.
For the opposite limit, $1-n_f \ll 1$,
we can develop a single-vacancy picture:
the missing $f$ moments act as scattering centers in the effective two-band
system, and the LDOS is now suppressed at the {\it vacancy locations}.
This picture is realized in Fig.~\ref{fig:ldos_nf} with $n_c\!=\!0.8$ fixed and varying $n_f$.
Surprisingly, a single-impurity behavior
(with {\it negative} correlation between LDOS and the site occupancy of $f$-moments)
is realized even at large $f$-moment concentrations like $n_f\!=\!0.5$.
It is only at $n_f=0.95$ that one finally sees the
single-vacancy behavior ({\it positive} correlation between
LDOS and site occupancy);
this is consistent with $n_f \approx n_c$ separating the HFL and
single-impurity FL.
Fig.~\ref{fig:ldos_nc} illustrates that varying $n_c$ can also be used
to tune between the single-impurity and single-vacancy limits
(here for a fixed realization of disorder!).

We now turn to the strongly inhomogeneous phases occurring for intermediate
values of $n_f$ around $n_f=n_c$.
Fig. ~\ref{fig:ldos_nc}b clearly shows island formation
in the low-energy LDOS.
This in turn gives rise to huge fluctuations in the Kondo temperatures of
individual moments, resulting in broad distributions of, e.g., $\chi_{\rm loc}$
and the existence of weakly screened moments down to lowest $T$.
Similar to impurities in disordered metals and
disordered Kondo lattices \cite{KondoDis},
these broad distributions of $T_K$ produce non-FL behavior in both thermodynamic and
transport properties -- these arguments
qualitatively apply to our situation as well.

\begin{figure}[t]
\centerline{\epsfig{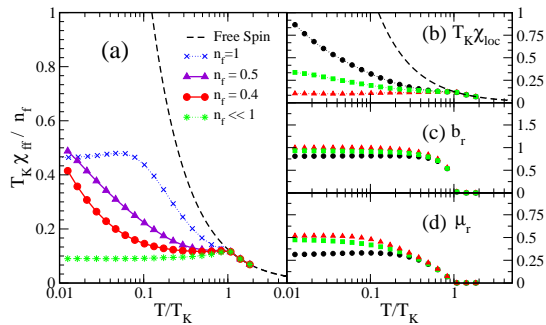}}
\caption{
(color online) Temperature dependence of $\chi$ and mean-field parameters.
(a) $f$-moment susceptibility $\chi_{ff}$;
in the inhomogeneous regime $n_f\approx n_c=0.4$, $\chi_{ff}$ does not saturate to the
lowest $T$ studied.
(b) Local susceptibility, $\chi_{\rm loc}(T)$, at three different sites, illustrating
the broad range of behavior.
(c,d) Mean-field parameters $b_r(T)$ and $\mu_r(T)$ at the same sites.
$J_K/D=1.25$ and for (b,c,d) $n_f=n_c=0.4$.
}
\label{fig:tdep}
\end{figure}

In Fig.~\ref{fig:tdep} we show numerical data illustrating the behavior of
the susceptibility; other observables (entropy, specific heat etc.) will be presented
elsewhere~\cite{TBP}.
Fig.~\ref{fig:tdep}(a) shows how $\chi_{ff}(T)$ in the Kondo lattice and impurity limits
saturate at low-$T$.
In contrast, for intermediate $n_f$ the $\chi_{ff}(T)$ does not saturate to the
lowest $T$ studied (finite-size effects prevent calculations at lower $T$).
While we cannot rule out saturation of $\chi_{ff}$ at extremely low temperatures,
it is clear that FL behavior is strongly suppressed in this regime,
as illustrated in the phase diagram, Fig.~\ref{fig:pd} --
this behavior is very reminiscent of the experimental data on \cecoin\ \cite{naka1}.
Fig.~\ref{fig:tdep}(b) shows $\chi_{\rm loc}(T)$ for three different sites,
demonstrating that $\chi_{\rm loc}$ is strongly inhomogeneous
(while the $b_r$, $\mu_r$ are more homogeneous),
and the behavior of $\chi_{ff}(T)$ arises from a fraction of
the $f$ moments only.

{\it Inter-moment correlations.}
So far, we have neglected the effects of a magnetic inter-moment interaction, $I$,
either due to direct exchange or mediated by the conduction electrons.
These interactions can be important for not too small $n_f$ \cite{fn1b}, provided
that their strength is comparable to or larger than $T_K$.
(Note that $I\gg T_K$ usually leads to a magnetically ordered state.)
In the dense limit, the coherence temperature will be enhanced compared
to the $I=0$ situation, as the
local moments are quenched by $I$ due to inter-moment correlations.
To assess the effect of $I$ for intermediate $n_f$ we have performed
calculations using the large-$N$ formalism of Ref.~\onlinecite{ffl}.
Our preliminary results (not shown) indicate that the rare unquenched moments
tend to be suppressed for larger $n_f$ and sufficiently strong $I$,
but islands and broad $\chi_{\rm loc}$ distributions survive for smaller $n_f$
over a sizable parameter range \cite{TBP}.
The resulting phase diagram is sketched in Fig.~\ref{fig:pd}b, and is qualitatively
remarkably similar to the one of \cecoin \cite{naka1}.


{\it Conclusions.}
We have studied randomly depleted Kondo lattices over the full
crossover from the dense lattice to the dilute impurity limit.
We found that $n_f=n_c$ appears to be the dividing line
between the two limiting FL behaviors;
in the vicinity of $n_f=n_c$ the system is generically strongly inhomogeneous and
displays non-FL behavior.
Our results are relevant to a variety of Kondo-lattice metals, e.g. \cecoin,
that show both HFL and single-impurity FL behavior as the $f$-moments are randomly depleted,
without signs of magnetic order.
While the condition $n_f\approx n_c$ likely only applies to heavy-fermion metals
described approximately by the $S=1/2$ single-channel model, Eq.~(\ref{KH}),
we expect the presence of two FL regimes separated by inhomogeneous phases
(Fig.~\ref{fig:pd}) to be a generic feature of $f$-moment-depleted
heavy FLs.
We speculate that a rather large (effective) $n_c$ is the reason why, e.g.,
\celapb\ displays single-impurity like behavior for sizeable range of $x\equiv n_f$.



We thank S. Burdin, S. Florens, A. Georges, A. Ghosal, A. Rosch,
J. Schmalian, and G. Zarand for discussions, and H. U. Baranger
for access to the computational facilities at Duke.
This research was supported by the
DFG Center for Functional Nano\-struc\-tures and the
Virtual Quantum Phase Transitions Institute (Karls\-ruhe),
by NSF Grants PHY99-0794 (KITP Santa Barbara), DMR-0506953 (RKK),
and the DAAD (RKK).



\begin{thebibliography}{}
\bibitem{fn1}
Consistent with the experimental data of Refs.~\onlinecite{naka1,celapb},
we assume upon random $f$-spin depletion the paramagnetic HFL does not
enter a magnetic state.
\bibitem{fn4}
The finite-$T$ phase transition is an artifact of mean-field theory;
however, the low-$T$ physics is known to be captured accurately by this approach.
\bibitem{fn3}
We expect the physics discussed here to be generic to
both two- and three-dimensional systems.
\end{thebibliography}
\end{document}